\documentclass{PoS}
\title{The Giant Radio Array for Neutrino Detection}

\ShortTitle{GRAND}

\author{Olivier Martineau-Huynh$^{1}$, Kumiko Kotera$^{2}$, Didier Charrier$^3$, Sijbrand De Jong$^4$, Krijn D. de Vries$^5$, Ke Fang$^6$, Zhaoyang Feng$^7$, Chad Finley$^8$, Quanbu Gou$^7$, Junhua Gu$^{9}$, Hongbo Hu$^7$,
 Kohta Murase$^{10}$, Valentin Niess$^{11}$, Foteini Oikonomou$^{10}$, Nicolas Renault-Tinacci$^{2}$, Julia Schmid$^{12}$, \speaker{Charles
Timmermans}$^{,3}$, Zhen Wang$^7$, Xiangping Wu$^{9}$, Jianli Zhang$^{9}$, Yi Zhang$^{7}$ \\
$^{1}$LPNHE, CNRS-IN2P3 and Universit\'es Paris VI \& VII, 4 place Jussieu, 75252 Paris, France\\ 
$^2$Institut d'Astrophysique de Paris,  Sorbonne Universit\'es, UPMC Univ. Paris 6 and CNRS, UMR 7095, 98 bis bd Arago, 75014 Paris, France\\
$^3$SUBATECH, IN2P3-CNRS, Universit\'e de Nantes, Ecole des Mines de Nantes, Nantes, France\\
$^4$Nikhef/Radboud University, Nijmegen, the Netherlands\\
$^5$Vrije Universiteit Brussel, Dienst ELEM, B-1050 Brussels, Belgium\\
$^6$Department of Astronomy and Astrophysics, University of Chicago, Chicago, IL 60637, USA\\
$^7$Key Laboratory of Particle Astrophysics, Institute of High Energy Physics, Chinese Academy of Sciences, Beijing 100049, China\\
$^8$Oskar Klein Centre and Dept. of Physics, Stockholm University, SE-10691 Stockholm, Sweden\\
$^9$National Astronomical Observatory, Chinese Academy of Sciences, Beijing 100012, China\\   
$^{10}$ Department of Physics, Department of Astronomy \& Astrophysics, Pennsylvania State University, University Park, PA, USA\\    
$^{11}$  Clermont Universit\'e, Universit\'e Blaise Pascal, CNRS/IN2P3, Laboratoire de Physique Corpusculaire, BP 10448, F-63000 Clermond-Ferrand, France \\
$^{12}$Laboratoire AIM, Universit\'e Paris Diderot/CEA-IRFU/CNRS, Service d'Astrophysique, CEA Saclay, 91191 Gif-sur-Yvette, France\\       
        E-mail: \email{omartino@in2p3.fr}, \email{kotera@iap.fr}
	}

\abstract{
High-energy neutrino astronomy will probe the working of the most violent phenomena in the Universe. The Giant Radio Array for Neutrino Detection (GRAND) project consists of an array of $\sim10^5$ radio antennas deployed 
over $\sim$\,200\,000\,km$^2$ in a mountainous site. It aims at detecting high-energy neutrinos via the measurement of air showers induced by the decay in the atmosphere of $\tau$ leptons produced by the interaction of the 
cosmic neutrinos under the Earth surface. Our objective with GRAND is to reach a neutrino sensitivity of $3\times10^{-11}E^{-2}$\,GeV$^{-1}$\,cm$^{-2}$\,s$^{-1}$\,sr$^{-1}$ above $3\times10^{16}$\,eV. This sensitivity 
ensures the detection of cosmogenic neutrinos in the most pessimistic source models, and about 100 events per year are expected for the standard models. GRAND would also probe the neutrino signals produced at the 
potential sources of UHECRs. 

We show how our preliminary design should enable us to reach our sensitivity goals, and present the  experimental characteristics. We assess the possibility to adapt GRAND to other astrophysical radio measurements. We discuss in this token the technological options for the detector and the steps to be taken to achieve the GRAND project.
}

\FullConference{The 34th International Cosmic Ray Conference,
		30 July- 6 August, 2015,
		The Hague, The Netherlands}

\begin{document}

Neutrinos are unique messengers that let us see deeper in objects, further in distance, and pinpoint the exact location of their production. These characteristics will help unveil the mystery behind the most violent 
astrophysical phenomena from which high-energy neutrinos should originate, and identify the source of their parent ultrahigh energy cosmic rays (UHECRs). 

The Giant Radio Array for Neutrino Detection (GRAND) aims primarily at reaching a sensitivity that ensures the detection of such neutrinos, even in the most pessimistic UHECR source models. For standard source models, it 
should detect enough events to launch neutrino astronomy.

The project consists of an array of $\sim 10^5$ radio antennas deployed over an area of $\sim$\,200\,000 \,km$^2$ in a mountainous site. GRAND will search for the radio signal emitted by the air showers of $\tau$ leptons 
produced by the interaction of cosmic neutrinos underground. 
We present the detection method in sec.~\ref{section:method}, the simulations that have led to the current design of GRAND in sec.~\ref{section:sensitivity}, and discuss in secs.~\ref{section:bg}-\ref{section:engineering} 
the steps to be taken to fully define our science case, elaborated in secs. \ref{section:sc}-\ref{section:cr}.

\section{Detection Method}\label{section:method}

Cosmic $\nu_{\tau}$s can produce $\tau$ particles underground through charged current interaction. $\tau$s travel to the surface of the Earth and decay in the atmosphere, generating Earth-skimming extensive air showers 
(EAS)\footnote{Other neutrino flavors can be neglected as the electron range in matter at these energies is too short and the muon decay length too large compared to flight distances underground.}~\cite{Fargion00,Bertou04}. Coherent electromagnetic radiation is associated to the shower development
at frequencies of a few to hundreds of MHz at a detectable level for showers with $E\gtrsim10^{17}$~eV. 

The strong beaming of the electromagnetic emission combined with the transparency of the atmosphere  to radio waves will allow the radio-detection of EAS initiated by $\tau$ decays at distances up to several 
tens of kilometers. Radio antennas are thus ideal instruments for this purpose. Furthermore, they offer practical advantages (limited unit cost, easiness of deployment, ...) 
that allow the deployment of an array over very large areas, as required by the expected low neutrino rate. 

Remote sites, with low electromagnetic background, should obviously be considered for the array location. In addition, mountain ranges are preferred, first because they offer an additional target for the neutrinos, and also because mountain slopes are better suited to the detection of horizontal showers compared to flat areas, parallel to the showers trajectories.

GRAND antennas are foreseen to operate in the $30-100$\,MHz frequency band. Short wave background prevents detection below this range, and above the coherence of the geomagnetic emission fades. 
However, an extension of the antenna response up to $200-300$\,MHz would enable us to observe the Cherenkov ring associated with the air shower, that represents a sizable fraction of the total electromagnetic 
signal and may provide an unambiguous signature for background rejection.

\section{GRAND layout and neutrino sensitivity}\label{section:sensitivity}
We present here a preliminary evaluation of the potential of GRAND for the detection of cosmic neutrinos. 
We initially simulated the response of a setup of 40\,000 antennas deployed on a square layout of 60\,000\,km$^2$ in a remote mountainous area (Tianshan mountains, XinJiang, China). 

{\bf Simulation method.}
We perform a 1D tracking of a primary $\nu_{\tau}$, simulated down to the converted $\tau$ decay. 
Standard rock with a density of 2.65~g/cm$^3$ 
is assumed down to sea level and the Earth core is modeled with the Preliminary Reference Earth Model \cite{Dziewonski81}. 
The simulation of the Deep Inelastic Scattering of the neutrinos is performed with Pythia6.4, using the CTEQ5d probability distribution functions (PDF) combined with \cite{Gandhi98} for cross section calculations. The propagation of the produced $\tau$ is simulated using randomized values from parameterisations of GEANT4.9 PDFs for 
$\tau$ path length and proper time. Photonuclear interactions in GEANT4.9 have been extended above PeV energies following \cite{Dutta00}. The $\tau$ decay is simulated using the TAUOLA package.

Recently EAS radio signals have been measured in good agreement with the predictions from several independent simulation codes \cite{Huege14}, but additional work is still required 
to properly model horizontal and upward-going trajectories. 
For the time being only an analytical parametrization of EAS radio detection has been performed, 
based on experimental results and simulation outputs. First, we assume that the EAS electromagnetic signal strength $\epsilon$ does not vary significantly along the shower axis for distances 
between 20 and 120~km from the $\tau$ decay point. This is supported by a simulation carried out with the SElFAS code~\cite{Marin12} for horizontal showers. 
The ANITA observation of UHECRs of energies around $10^{19}$~eV at distances beyond 100~km provides experimental support for this hypothesis \cite{Hoover10}.
Next we consider that the field strength scales linearly with energy \cite{Nelles15}, and that the lateral profile drops exponentially $\epsilon(\alpha)=KE\exp(-{L \tan \alpha}/{\lambda})$, where $\alpha$ is 
the angle under which the shower maximum is seen from the antenna position with respect to the shower axis, $K$ a linear coefficient, $\lambda$ the lateral profile exponential attenuation parameter and 
$L$ the projection on the shower axis of the distance between the antenna and shower maximum. 
We define a detection threshold angle $\alpha_{\rm th}$, below which the field is strong enough for detection. Experimental results\footnote{Vertical showers of energy $\sim 10^{17}$\,eV are detected at
distances up to $\sim$300~m from shower axis \cite{Nelles15,Rebai12}.} indicate a value $\alpha_{\rm th} \sim 3^\circ$ for $E=10^{17}$~eV. We can then write $\tan(\alpha_{\rm th})=\frac{\lambda}{L} \log({E}/{10^{17}}) + 0.05$.

We consider that a shower is detected if its energy exceeds a value $E_{\rm th}$ and if a cluster of 8 antennas or more are in direct view of the $\tau$ decay point, at a distance between 20 and 120~ km,
and inside a cone of half angle $\alpha_{\rm th}$ around the shower axis. We evaluate two different approaches: conservative ($E_{\rm th}=10^{17}$~eV, $\lambda=200$~m, yielding $\alpha_{\rm th}=7^\circ$ at $10^{18}$~eV and $12^\circ$ at $10^{19}$~eV) and aggressive ($E_{\rm th}=3\times10^{16}$~eV, $\lambda=400$~m, yielding $\alpha_{\rm th}=12^\circ$ at $10^{18}$~eV and $20^\circ$ at $10^{19}$~eV). 

Note that the Cherenkov effect is not taken into account in this study.
This will be done in the up-coming full Monte-Carlo study of the GRAND neutrino sensitivity.

{\bf Results and implications.}
Assuming a 3-year observation on an area of 60\,000 km$^2$ with no neutrino candidate, a 90\%\,C.L. integral limit of $6.6 \,(3.1)\times10^{-10}$ GeV$^{-1}$~cm$^{-2}$~s$^{-1}$ can be derived for an $E ^{-2}$ neutrino flux in our conservative (aggressive) scenario.
 This is a factor $\ge 5$ better sensitivity than that of other projected giant neutrino telescopes for EeV energies \cite{Barwick11}.

\begin{figure}[t]
\begin{center}
{\includegraphics[width=0.35\columnwidth]{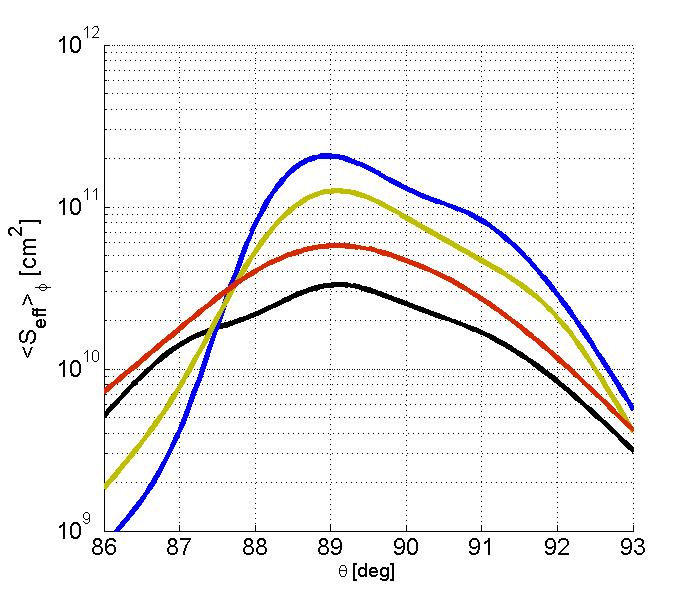}}
{\includegraphics[width=0.4\columnwidth]{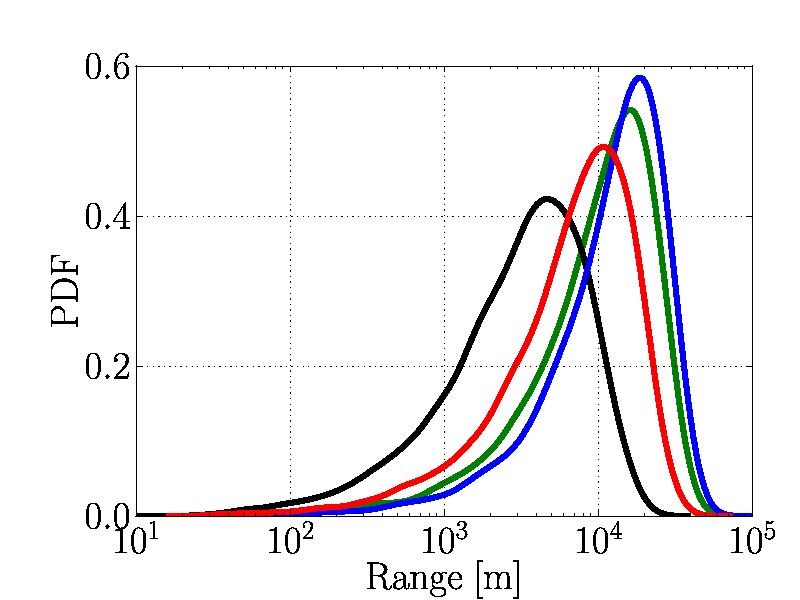}}
\end{center}
\caption{{\it Left:} Effective area of the 60\,000 km$^2$ simulated set-up as a function of zenith angle $\theta$ for various initial $\nu_{\tau}$ energies. 
Zenith angles $<90^\circ$ correspond to down-going trajectories. 
 {\it Right:} Flight distance probability density functions for $\tau$s produced by the interaction of $\nu_{\tau}$ of various initial energies in a 30~km thick wall.
 For both plots black: $10^{17.5}$~eV, red: $10^{18.5}$~eV, green: $10^{19.5}$~eV, blue:$10^{20.5}$~eV).}
\label{fig:Sim}
\end{figure}

This preliminary analysis also demonstrates that mountains constitute a sizable target for neutrinos, with $\sim$50\% of down-going events, corresponding to neutrinos interacting inside the mountains (Fig.~\ref{fig:Sim}). 
It also appears that specific parts of the array (large mountains slopes facing another mountain range at distances of $30-80$\,km) are associated with a detection rate well above the average. 
We thus realized that a factor of 10 improvement in sensitivity -- corresponding to cosmogenic neutrinos event rates between 10 to 100 per year -- could be reached with a factor of 3 increase in the detector area, 
provided the detector is composed of several sub-arrays of smaller size (few 10\,000~km$^2$) deployed solely on favorable sites. This is the envisioned GRAND setup.

The expected angular resolution on the arrival direction of the events detected was computed analytically following \cite{Ardouin11}. A mean resolution of 0.05$^\circ$ 
should be achievable for a 3~ns precision on the antenna trigger timing. Energy estimation will be challenging, mainly because the $\tau$ decay point is unknown.
However, the $\nu_{\tau}$ energy is correlated to measurable parameters (Fig.~\ref{fig:Sim}) which could provide a handle to statistically create the neutrino energy spectrum.

\section{Background rejection}\label{section:bg}
A few to a hundred cosmic neutrinos per year are expected in GRAND. The rejection of events initiated by high energy particles other than cosmic neutrinos should be manageable, as: 
i) the flux of atmospheric neutrinos is negligible above $10^{16}$~eV \cite{Aartsen14b}, ii) the rate of showers generated by muon decay above GRAND is expected to be few per century \cite{Chirkin04}, and 
iii) rejection of all trajectories reconstructed down to $1^\circ$ below the horizon would strongly suppress the measured rate of EAS initiated by standard cosmic rays, without significantly affecting 
GRAND neutrino sensitivity.

The event rates associated to terrestrial sources (human activities, thunderstorms, ...) are difficult to evaluate, but a conservative estimate can be derived from the results of the Tianshan Radio Experiment for Neutrino
Detection (TREND). TREND~\cite{Ardouin11} is an array of 50 self-triggered antennas deployed in a populated valley of the Tianshan mountains, with antenna design and sensitivity similar to what is
foreseen for GRAND. The observed rate of events triggering antennas over a surface $\gtrsim
1$\,km$^2$ was 15 events/day in TREND, which scales to a safe estimate of $\sim10^9$ events/year for a 200\,000\,km$^2$ array. A background rejection rate better than $10^9$ is therefore probably necessary to reach the 
expected sensitivity of GRAND. Meeting this requirement without affecting the neutrino detection efficiency is a key challenge for the GRAND project. 
 
Amplitude patterns on the ground (emission beamed along the shower axis or signal enhancement along the Cherenkov ring\footnote{{https://indico.in2p3.fr/event/10976/session/3/contribution/40/material/slides/0.pdf}}), as well as wave polarization \cite{Aab14} are strong signatures of 
$\nu$-initiated air showers that could provide efficient discrimination tools. These options are being investigated within GRAND, through simulations and experimental work. In 2016 the GRANDproto project \cite{Gou15} 
in particular, will deploy an hybrid detector composed of 35 3-arm antennas (allowing a complete measurement of the wave polarization) and 24 scintillators, that will cross-check the EAS nature of radio-events selected from a polarization signature compatible with EAS.

A project is also being initiated to perform similar tests at Auger-AERA \cite{Schulz15}. Its goal is to make use of the good detection efficiency of the Auger fluorescence detectors down to large zenith angles
to complement the GRANDproto study with very inclined EAS trajectories.

\section{GRAND engineering array}\label{section:engineering}
Detailed design of the GRAND detector will be refined once the technical requests associated to the definite science case of GRAND have been defined. It is however certain that the GRAND detector will be a technical 
challenge given its unprecedented size. Its huge number of units necessarily implies that they should be fully autonomous over periods of years, with a very low failure rate. The driving principles of GRAND to ensure its 
success will be to stay as modular and basic as possible and rely on standard and validated industrial-scale solutions whenever possible.
Before considering the complete GRAND layout, an engineering array of size $\sim 1000$\,km$^2$ will be deployed in order to test the proposed technological solutions. This array will obviously be too small to perform a
neutrino search, but cosmic rays should be detected above $10^{18}\,$eV. Their reconstructed properties (energy spectrum, directions of arrival, nature of the primary) will enable us to validate this stage, if found to be 
compatible with the expectations. The absence of events below the horizon will also test our EAS identification strategy.

\section{High-energy neutrino astronomy with GRAND}\label{section:sc}
\begin{figure}[t]
\begin{center}
{\includegraphics[width=0.48\columnwidth]{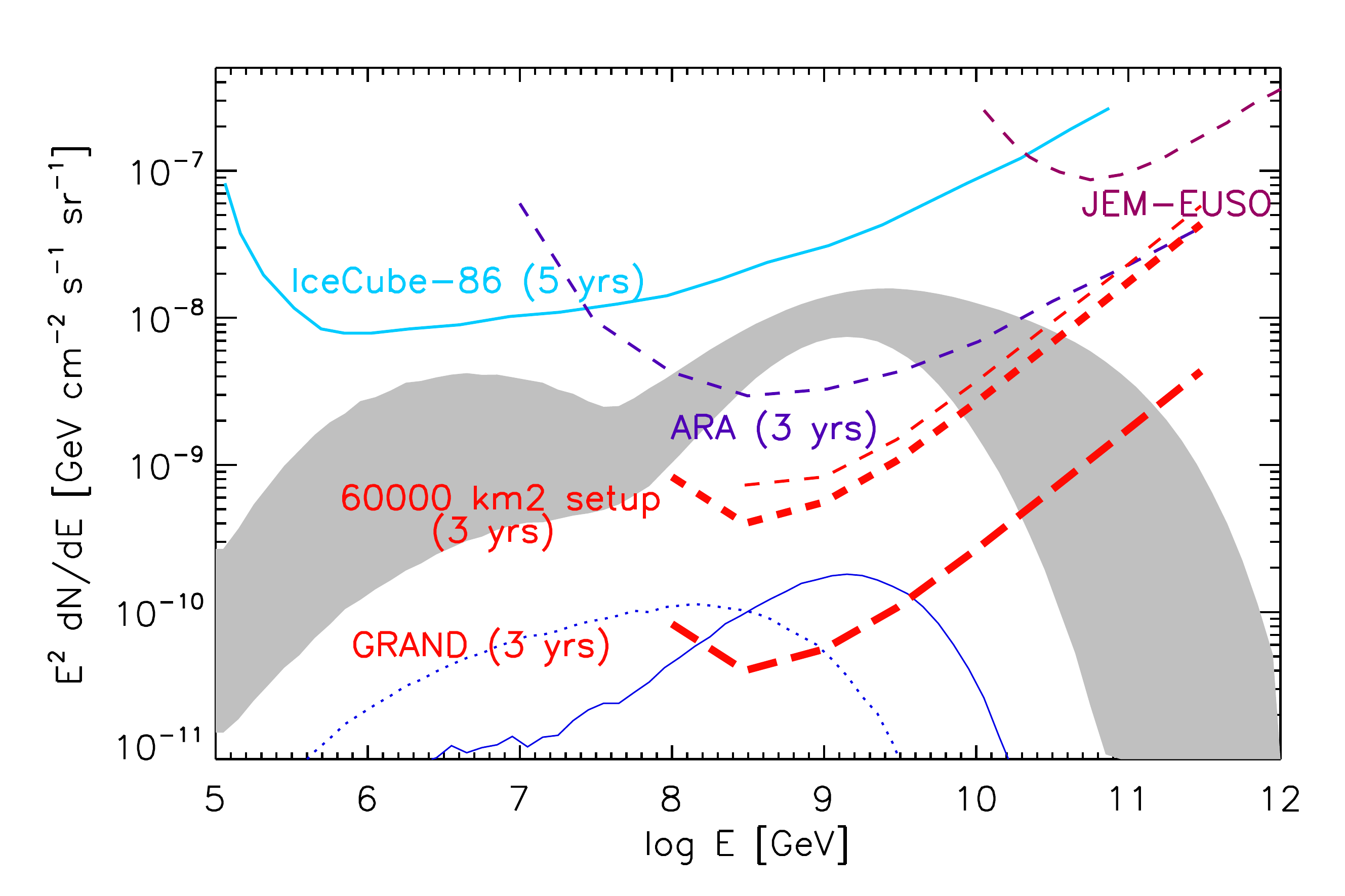}}
{\includegraphics[width=0.48\textwidth]{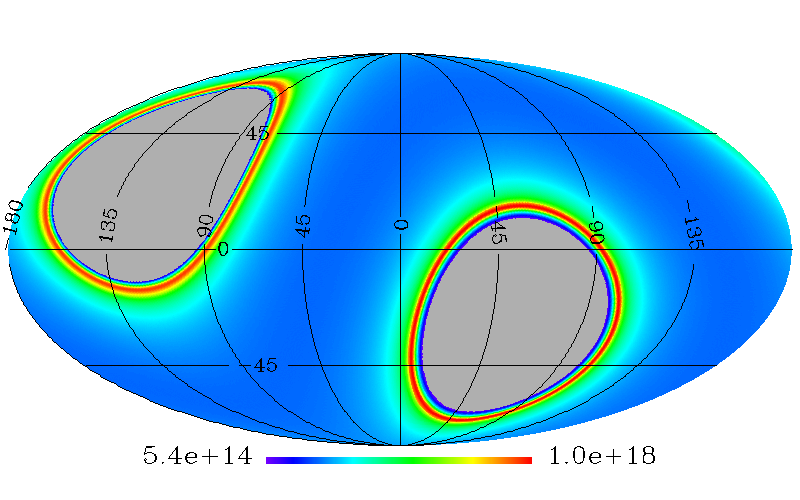}}
\end{center}
\caption{{\it Left:} Differential sensitivity limit of the 60\,000 km$^2$ simulated setup (red dashed lines,  thin: conservative, thick: aggressive), of the projected 3 times larger GRAND array (red thick long-dashed), and for other instruments, and estimated theoretical cosmogenic neutrino fluxes for all flavors \cite{KAO10}. The blue line gives the most pessimistic fluxes (pure iron UHECR injection and low maximum acceleration), while the gray shaded-region indicates the ``reasonable'' parameter range. {\it Right:} Integrated exposure of the simulated setup over 3 years at $10^{18.5}\,$eV, in units of cm$^2$\,s, in Galactic coordinates.}
\label{fig:SC}
\end{figure}

UHECRs are likely produced in extragalactic sources, given the strength of Galactic magnetic fields and the lack of correlations with the Galactic plane. Some fraction of their energy is converted to high-energy neutrinos through the decay of charged pions produced by interactions with ambient matter and radiation. This can happen in the source environment or during the flight in the intergalactic medium ({\it cosmogenic} neutrinos). 
The range of expected cosmogenic neutrino fluxes can be calculated precisely, and depends mostly on parameters inherent to the cosmic rays (Fig.~\ref{fig:SC}) \cite{KAO10}. The calculated level of high-energy neutrino fluxes produced at the sources depends on the modeling of the acceleration region. But robust flux estimates have recently been derived for some sources~\cite{Baerwald15,Murase09,FKMO14,Murase14}, arguing that many scenarios can be tested with increased sensitivities. 

The sensitivity of GRAND should guarantee the detection of cosmogenic EeV neutrinos. For reasonable source scenarios (Fig.~\ref{fig:SC}), GRAND aims at collecting in the order of $100$ events per year above 
$3 \times 10^{16}\,$eV, which helps to understand the underlying components. An energy resolution of 50\% would enable us to picture the shape of the diffuse energy spectrum, but the most important information would stem from the precise ($<0.1^\circ$) angular resolution on the arrival directions. 
Cross-correlating the position of sources and of events, it will be possible to discriminate between cosmogenic neutrinos and those produced in the source environment. The former should indeed correlate with the large-scale structures at high redshift where the integrated flux is maximal.  

The ideal way to identify high-energy neutrino sources would be to observe a point source ---a {\it transient} source is a likely candidate due to the power and density constraints from UHECR observations (e.g.,
\cite{Murase_Takami09}). GRAND opens this possibility with its excellent spatial resolution and its sky coverage. In the simulation layout, the instantaneous field of view of GRAND is a band corresponding to zenith angles
$80^\circ\leq\theta\leq100^\circ$, and its integrated exposure $\sim 10^{16}\,$cm$^2$\,s for 3 years in the energy range $\sim10^{17-20}\,$eV, over a large portion of the sky (Fig.~\ref{fig:SC}). Short-lived (< day) transients have a low probability of being spotted, but for longer 
transients (blazar flares, ultraluminous supernovae, magnetars etc.), post-analysis at the location of existing transients and stacking searches can be done efficiently. Depending on the background discrimination efficiency, GRAND would be able to send alerts to other experiments via AMON-type structures \cite{Smith13}. 

The design of GRAND is still preliminary, and can be adapted to accommodate further requests related to the high-energy neutrino astronomy. The large array size and the frequency range of GRAND -- if extended to 300\,MHz -- could be profitable for 
other physics goals (such as the study of the Epoch of Reionization \cite{Zaroubi13} or the search for Fast Radio Bursts \cite{Kulkarni14}). We are presently assessing the technical cost of pushing the detector towards these directions. 

\section{Ultrahigh Energy Cosmic Rays}\label{section:cr}
The energy spectrum of cosmic rays shows an increased steepening at around $10^{19.6}$\,eV
\cite{Auger13,Tinyakov14}. The origin of this steepening is still unclear.
The Telescope Array (TA) analysis is consistent with a dominance of protons up to the highest energies~\cite{Tinyakov14}, while the Auger experiment favours a mixed composition~\cite{Auger14}.
Also TA has reported an anisotropy in the arrival directions of UHECR in the Northern hemisphere~\cite{TA14}, whereas the Auger collaboration has not 
identified a significant anisotropy at the highest energies in the South \cite{Auger15}. 
Due to the absence of photons at high energies, top-down and other exotic scenarios of UHECR production are disfavoured.

Given its effective area, GRAND would be the largest UHECR observatory on ground, and should provide valuable information to help resolve these puzzles if precision of reconstructed parameters 
is satisfactory. Determination of the UHECR composition in particular is a key issue. Detailed simulation studies will be performed to estimate performances reachable by GRAND.

\section{Conclusion}
The GRAND project aims at building a next-generation neutrino telescope composed of radio antennas deployed over $\sim$\,200\,000\,km$^2$. Preliminary simulations indicate that
 a sensitivity guaranteeing the detection of cosmogenic neutrinos could be achievable. GRAND would also be a powerful instrument for UHECR detection at energy $\gtrsim 10^{19.6}\,$eV with high statistics, 
 and possibly for other science objectives (Epoch of Reionzation, Fast Radio Bursts). Work is ongoing to determine precisely the achievable scientific goals and the corresponding technical constraints. 
 Background rejection strategies and technological options are also being investigated.\\

\noindent\footnotesize{{\it Acknowledgements.} We thank J. Alvarez-Muniz and W. Carvalho for helpful comments and
discussions. The GRAND and GRANDproto projects are supported by the Institut Lagrange de Paris, the France China Particle Physics
Laboratory, the Natural Science Fundation of China (Nos.11135010, 11375209) and the Chinese Ministery of Science and Technology.}

\bibliography{ICRC}

\providecommand{\href}[2]{#2}\begingroup\raggedright\begin{thebibliography}{10}

\bibitem{Fargion00}
D.~{Fargion}, {\it {Horizontal Tau air showers from mountains in deep valley:
  Traces of Ultrahigh neutrino tau}},  {\em International Cosmic Ray
  Conference} {\bf 2} (Aug., 1999) 396,
  [\href{http://arxiv.org/abs/astro-ph/9906450}{{\tt astro-ph/9906450}}].

\bibitem{Bertou04}
X.~{Bertou} and et~al, {\it {Tau neutrinos in the Auger Observatory: a new
  window to UHECR sources}},  {\em Astroparticle Physics} {\bf 17} (May, 2002)
  183--193, [\href{http://arxiv.org/abs/astro-ph/0104452}{{\tt
  astro-ph/0104452}}].

\bibitem{Dziewonski81}
A.~{Dziewonski} and D.~{Anderson}, {\it {Preliminary reference Earth model}},
  {\em Physics of the Earth and Planetary Interiors} {\bf 25} (June, 1981)
  297--336.

\bibitem{Gandhi98}
R.~{Gandhi}, C.~{Quigg}, M.~H. {Reno}, and I.~{Sarcevic}, {\it {Neutrino
  interactions at ultrahigh energies}},  {\em Phys. Rev. D} {\bf 58} (Nov.,
  1998) 093009, [\href{http://arxiv.org/abs/hep-ph/9807264}{{\tt
  hep-ph/9807264}}].

\bibitem{Dutta00}
S.~{Dutta} and et~al., {\it {Searches for the appearance of {$\nu$} $_{tau }$
  using neutrino beams from muon storage rings}},  {\em European Physical
  Journal C} {\bf 18} (Dec., 2000) 405--416,
  [\href{http://arxiv.org/abs/hep-ph/9905475}{{\tt hep-ph/9905475}}].

\bibitem{Huege14}
T.~{Huege}, {\it {The Renaissance of Radio Detection of Cosmic Rays}},  {\em
  Brazilian Journal of Physics} {\bf 44} (Oct., 2014) 520--529,
  [\href{http://arxiv.org/abs/1310.6927}{{\tt arXiv:1310.6927}}].

\bibitem{Marin12}
V.~Marin and B.~Revenu, {\it {Coherent radio emission from cosmic ray air
  showers computed by Monte-Carlo simulation with SElFAS}},  {\em
  Nucl.Instrum.Meth.} {\bf A662} (2012) S171--S174.

\bibitem{Hoover10}
S.~{Hoover} and et~al., {\it {Observation of Ultrahigh-Energy Cosmic Rays with
  the ANITA Balloon-Borne Radio Interferometer}},  {\em Physical Review
  Letters} {\bf 105} (Oct., 2010) 151101,
  [\href{http://arxiv.org/abs/1005.0035}{{\tt arXiv:1005.0035}}].

\bibitem{Nelles15}
A.~{Nelles}, S.~{Buitink}, H.~{Falcke}, J.~R. {H{\"o}randel}, T.~{Huege}, and
  P.~{Schellart}, {\it {A parameterization for the radio emission of air
  showers as predicted by CoREAS simulations and applied to LOFAR
  measurements}},  {\em Astroparticle Physics} {\bf 60} (Jan., 2015) 13--24,
  [\href{http://arxiv.org/abs/1402.2872}{{\tt arXiv:1402.2872}}].

\bibitem{Rebai12}
A.~{Rebai}, P.~{Lautridou}, A.~{Lecacheux}, and O.~{Ravel}, {\it {Correlations
  in energy in cosmic ray air showers radio-detected by CODALEMA}},  {\em ArXiv
  e-prints} (Oct., 2012) [\href{http://arxiv.org/abs/1210.1739}{{\tt
  arXiv:1210.1739}}].

\bibitem{Barwick11}
S.~{Barwick}, {\it {ARIANNA - A New Concept for High Energy Neutrino
  Detection}},  {\em International Cosmic Ray Conference} {\bf 4} (2011) 238.

\bibitem{Ardouin11}
D.~{Ardouin} et~al., {\it {First detection of extensive air showers by the
  TREND self-triggering radio experiment}},  {\em Astroparticle Physics} {\bf
  34} (Apr., 2011) 717--731, [\href{http://arxiv.org/abs/1007.4359}{{\tt
  arXiv:1007.4359}}].

\bibitem{Aartsen14b}
M.~G. {Aartsen} et~al., {\it {Search for a diffuse flux of astrophysical muon
  neutrinos with the IceCube 59-string configuration}},  {\em Phys. Rev. D}
  {\bf 89} (Mar., 2014) 062007.

\bibitem{Chirkin04}
D.~{Chirkin}, {\it {Fluxes of Atmospheric Leptons at 600 GeV - 60 TeV}},  {\em
  ArXiv High Energy Physics - Phenomenology e-prints} (July, 2004)
  [\href{http://arxiv.org/abs/hep-ph/0407078}{{\tt hep-ph/0407078}}].

\bibitem{Aab14}
A.~{Aab} et~al., {\it {Probing the radio emission from air showers with
  polarization measurements}},  {\em Phys. Rev. D} {\bf 89} (Mar., 2014)
  052002, [\href{http://arxiv.org/abs/1402.3677}{{\tt arXiv:1402.3677}}].

\bibitem{Gou15}
Q.~{Gou}, {\it {R\&D on EAS radio detection with GRANDproto}},  {\em these
  proceedings} (2015).

\bibitem{Schulz15}
J.~{Schulz}, {\it {Status and Prospects of the Auger Engineering Radio Array}},
   {\em these proceedings} (2015).

\bibitem{KAO10}
K.~{Kotera}, D.~{Allard}, and A.~V. {Olinto}, {\it {Cosmogenic neutrinos:
  parameter space and detectabilty from PeV to ZeV}},  {\em J. Cos. and Astro.
  Phys.} {\bf 10} (Oct., 2010) 13--37,
  [\href{http://arxiv.org/abs/1009.1382}{{\tt arXiv:1009.1382}}].

\bibitem{Baerwald15}
P.~{Baerwald}, M.~{Bustamante}, and W.~{Winter}, {\it {Are gamma-ray bursts the
  sources of ultra-high energy cosmic rays?}},  {\em Astroparticle Physics}
  {\bf 62} (Mar., 2015) 66--91, [\href{http://arxiv.org/abs/1401.1820}{{\tt
  arXiv:1401.1820}}].

\bibitem{Murase09}
K.~{Murase}, P.~{M{\'e}sz{\'a}ros}, and B.~{Zhang}, {\it {Probing the birth of
  fast rotating magnetars through high-energy neutrinos}},  {\em Phys.~Rev.~D}
  {\bf 79} (May, 2009) 103001--103006,
  [\href{http://arxiv.org/abs/0904.2509}{{\tt arXiv:0904.2509}}].

\bibitem{FKMO14}
K.~{Fang}, K.~{Kotera}, K.~{Murase}, and A.~V. {Olinto}, {\it {Testing the
  newborn pulsar origin of ultrahigh energy cosmic rays with EeV neutrinos}},
  {\em Phys. Rev. D} {\bf 90} (Nov., 2014) 103005.

\bibitem{Murase14}
K.~{Murase}, Y.~{Inoue}, and C.~D. {Dermer}, {\it {Diffuse neutrino intensity
  from the inner jets of active galactic nuclei: Impacts of external photon
  fields and the blazar sequence}},  {\em Phys. Rev. D} {\bf 90} (July, 2014)
  023007, [\href{http://arxiv.org/abs/1403.4089}{{\tt arXiv:1403.4089}}].

\bibitem{Murase_Takami09}
K.~{Murase} and H.~{Takami}, {\it {Implications of Ultra-High-Energy Cosmic
  Rays for Transient Sources in the Auger Era}},  {\em ApJ Letters} {\bf 690}
  (Jan., 2009) L14--L17, [\href{http://arxiv.org/abs/0810.1813}{{\tt
  arXiv:0810.1813}}].

\bibitem{Smith13}
M.~W.~E. {Smith} et~al., {\it {The Astrophysical Multimessenger Observatory
  Network (AMON)}},  {\em Astroparticle Physics} {\bf 45} (May, 2013) 56--70,
  [\href{http://arxiv.org/abs/1211.5602}{{\tt arXiv:1211.5602}}].

\bibitem{Zaroubi13}
S.~{Zaroubi}, {\it {The Epoch of Reionization}},  in {\em Astrophysics and
  Space Science Library} (T.~{Wiklind}, B.~{Mobasher}, and V.~{Bromm}, eds.),
  vol.~396 of {\em Astrophysics and Space Science Library}, p.~45, 2013.

\bibitem{Kulkarni14}
S.~R. {Kulkarni}, E.~O. {Ofek}, J.~D. {Neill}, Z.~{Zheng}, and M.~{Juric}, {\it
  {Giant Sparks at Cosmological Distances?}},  {\em The Astrophysical Journal}
  {\bf 797} (Dec., 2014) 70, [\href{http://arxiv.org/abs/1402.4766}{{\tt
  arXiv:1402.4766}}].

\bibitem{Auger13}
{The Pierre Auger Collaboration}, {\it {The Pierre Auger Observatory:
  Contributions to the 33rd International Cosmic Ray Conference (ICRC 2013)}},
  \href{http://arxiv.org/abs/1307.5059}{{\tt arXiv:1307.5059}}.

\bibitem{Tinyakov14}
{The Telescope Array collaboration}, {\it {Latest results from the telescope
  array}},  {\em Proceedings of 4th Roma ICAP NIM-A} {\bf 742} (2014) 29.

\bibitem{Auger14}
{The Pierre Auger collaboration}, {\it {Depth of maximum of air-shower profiles
  at the Pierre Auger Observatory. II Composition implications}},  {\em Phys.
  Rev. D} {\bf 90} (2014) 122006.

\bibitem{TA14}
{The Telescope Array collaboration}, {\it {Indications of Intermediate-Scale
  Anisotropy of Cosmic Rays with Energy Greater than 57 EeV in the Northern Sky
  Measured with the Surface Detector of the Telescope Array Experiment}},  {\em
  ApJ letters 790} {\bf 790} (2014) L21.

\bibitem{Auger15}
{The Pierre Auger collaboration}, {\it {Searches for Anisotropies in the
  Arrival Directions of the Highest Energy Cosmic Rays Detected by the Pierre
  Auger Observatory}},  {\em ApJ} {\bf 804} (2015) 15.

\end{thebibliography}\endgroup
\end{document}